\documentclass[a4paper]{jpconf}
\usepackage{graphicx}
\usepackage{color}
\begin{document}

\title{Search for the time evolution of the synchrotron X-ray spectrum of the youngest Galactic supernova remnant G1.9+0.3 using {\it Suzaku}}

\author{A Sezer$^{1}$, R Yamazaki$^{2}$, Y Ohira$^{2}$, S Tanaka$^{3}$ and S Kisaka$^{2}$}

\address{$^1$ Department of Electrical-Electronics Engineering, Avrasya University, 61250, Trabzon, Turkey}
\address{$^2$ Department of Physics and Mathematics, Aoyama Gakuin University, 5-10-1 Fuchinobe, Sagamihara 252-5258, Japan}
\address{$^3$ Department of Physics, Faculty of Science and Engineering, Konan University, 8-9-1 Okamoto, Kobe, Hyogo 658-8501, Japan}

\ead{aytap.sezer@avrasya.edu.tr}

\begin{abstract}
G1.9+0.3 is the youngest known Galactic supernova remnant (SNR) and dominated by X-ray synchrotron emission. Synchrotron X-rays can be a useful tool to study the electron acceleration in young SNRs. The X-ray spectra of young SNRs give us information about the particle acceleration at the early stages of evolution of SNRs. In this work, we investigate the time evolution of roll-off frequency of the synchrotron spectrum from SNR G1.9+0.3 using {\it Suzaku}. For this analysis, we use $\sim$101 ks (2011) and $\sim$92 ks (2015) observations with the X-ray Imaging Spectrometer. We find that there are no significant differences in the spectral parameters and interpret our results.

\end{abstract}

\section{Introduction}

The X-ray spectra of several young SNRs are dominated by non-thermal emission: e.g., SN 1006 [1], G347.3$-$0.5 [2, 3] and G266.2$-$1.2 [4]. These discoveries in shell-like SNRs suggest that relativistic electrons are accelerated by SNR shocks. Their spectra have a power-law form (the synchrotron flux density at frequency ${\nu}$ is $F_{\nu} \propto \nu^{- \alpha} \propto \nu^{1 - \Gamma_{\rm x}}$, where ${\alpha}$ is the spectral index and ${\Gamma_{\rm x}}$ is the photon index). {\it Chandra} X-ray spectra showed that G1.9$+$0.3 is a young and X-ray synchrotron-dominated shell type SNR [5-7]. The age of this youngest known Galactic supernova remnant is estimated to be $\sim$110 yr [6]. Using {\it Chandra} and {\it NuStar} data, Zoglauer et al. [8], performed a detailed X-ray study of the SNR and found a spectral index of $\alpha$ $\sim$ 0.63, and a roll-off frequency of $\nu_{\rm rolloff}$ $\sim$$ 3.1\times10^{17}$ Hz. From {\it Suzaku} 2011 observation, G\"{o}k and Ergin [9], obtained a spectral index of 􏰃$\sim$0.61 and a roll-off frequency of 􏰃$\sim$ $3.1\times10^{17}$ Hz. 

The $\nu$$F_{\nu}$ spectrum of synchrotron X-ray emission has a peak around the roll-off frequency, $\nu_{\rm roll}$, which is related to the maximum electron energy of accelerated electrons ($E_{\rm max, e}$) and the magnetic field ($B$), $\nu_{\rm roll}$ $\propto$ $E_{\rm max, e}^2$ $B$ [10, 11].  The value of $\nu_{\rm roll}$ decreases as SNRs evolve (e.g., [12]). The SNR G1.9+0.3 is one of the best studied SNRs to search for the time variability. In this work, we investigate the time evolution of roll-off frequency of the synchrotron spectrum from G1.9+0.3. For this study, we use 2011 and 2015 observations with the X-ray Imaging Spectrometer (XIS; [13]) on board {\it Suzaku} [14], which has a high spectral resolution. In Section 2, we describe the {\it Suzaku} XIS observations and data reduction.  The spectral analysis is presented in Section 3. Finally, in Section 4, we discuss our results.

%% X-ray Observation and Data Reduction
\section{Observations and Data Reduction}

SNR G1.9+0.3 observed with XIS on 2011 March (Obs ID: 505053010) and 2015 March (Obs ID:509003010) for $\sim$101 ks and $\sim$92 ks, respectively. A detailed spectral analysis of {\it Suzaku} observation in 2011 was previously presented in [9]. The XIS consists of four X-ray charge-coupled devices (CCDs). Three of them (XIS0, 2, and 3) are front-illuminated CCDs, and the other (XIS1) is a back-illuminated CCD. After 2006 November 9, the XIS2 was out of operation. Therefore, we use only XIS0, 1, and 3 data.

We retrieved the data from the public {\it Suzaku} science data archive through the DARTS interface\footnote{https://darts.isas.jaxa.jp/astro/suzaku/}. Data reduction and analysis were made using {\sc HEAsoft} package\footnote{https://heasarc.nasa.gov/lheasoft/} version 6.20, {\sc xspec} version 12.9.1 and AtomDB version 3.0.8. The redistribution matrix file and ancillary response file were produced by {\sc xisrmfgen} and {\sc xissimarfgen} [15], respectively.

% X-ray Spectral Analysis
\section{Spectral Analysis} \label{Spectral analysis}

Figure 1 shows the {\it Suzaku} XIS images of 2011 and 2015 observations in the 0.3$-$10.0 keV. In order to investigate the time variability of X-ray emission from G1.9+0.3, we extracted XIS spectra from a circular region at the source with a radius of 3.1 arcmin centered on the remnant for both data sets. To estimate the background, we extracted data from source-free region on the same field of view. We subtracted the non-X-ray background (NXB) from both observations. The NXB spectra were made by using {\sc xisnxbgen} [16].  

\begin{figure}
\centering \vspace*{1pt}
\includegraphics[width=0.40\textwidth]{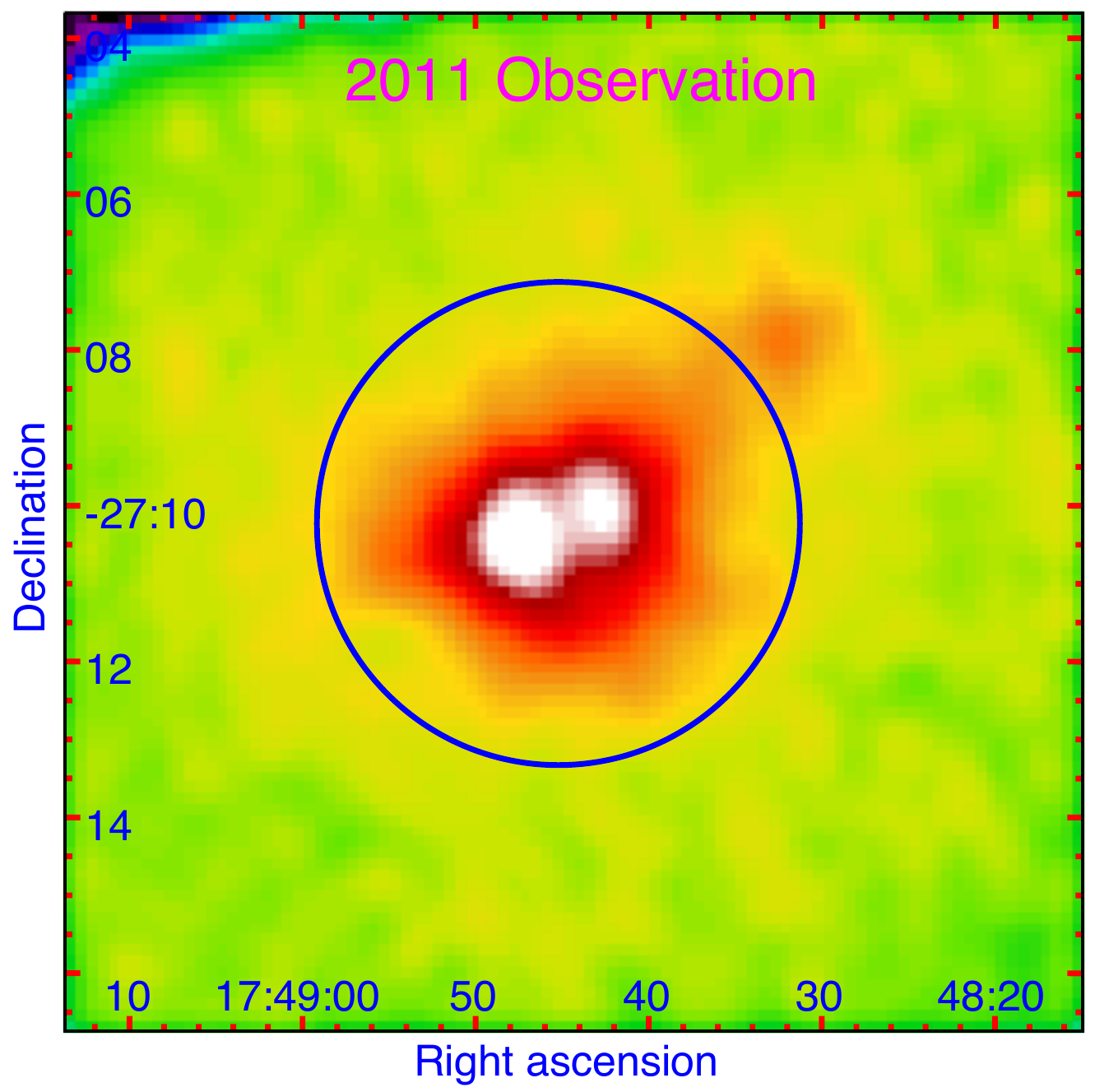}
\includegraphics[width=0.40\textwidth]{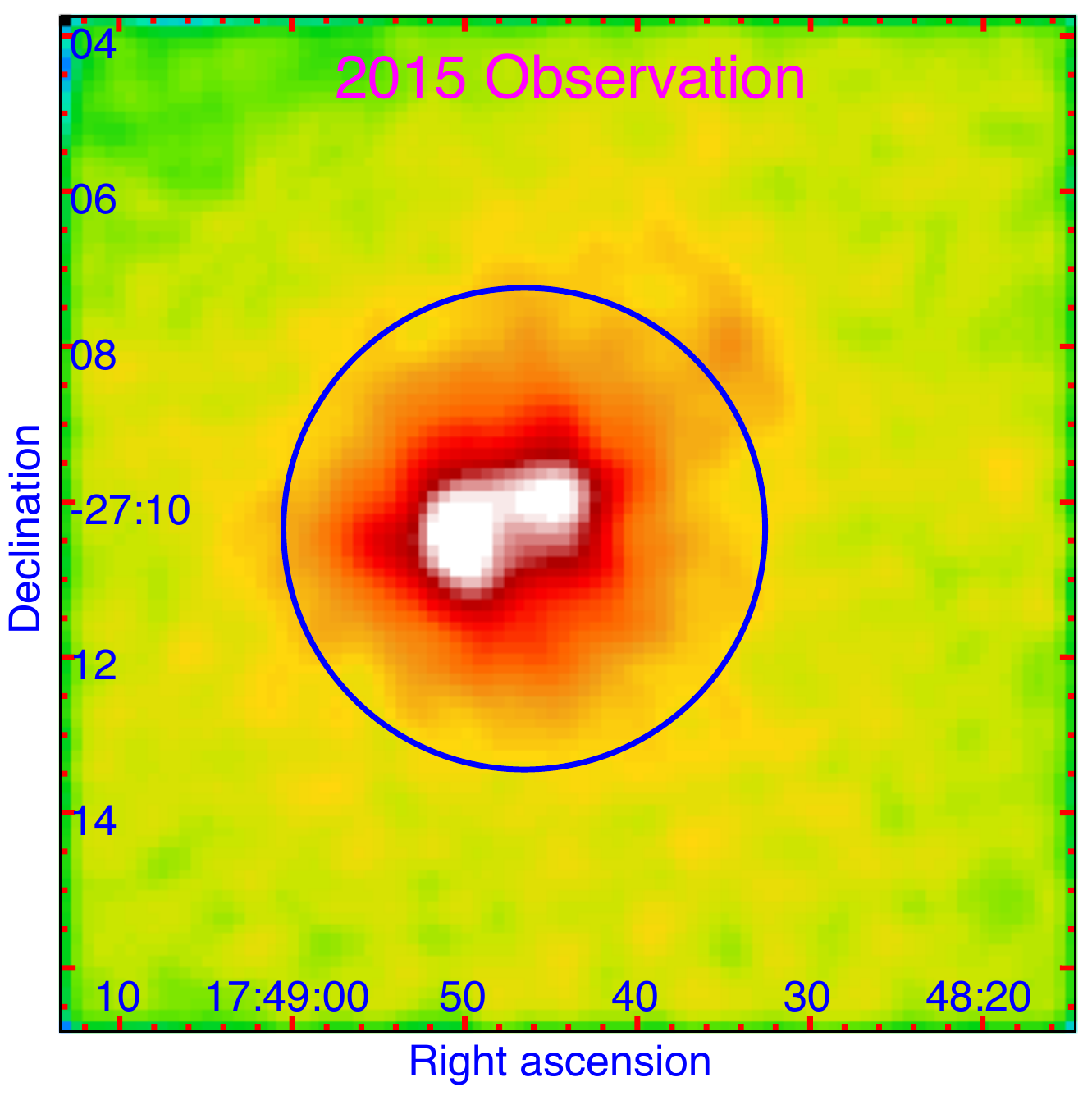}
\caption{XIS images of G1.9+0.3 in the 0.3$-$10.0 keV energy band. The circles indicate the extraction regions used for the spectral analysis.}
\label{figure_1}
\end{figure}

First, we fitted both spectra with a power-law (PL) model modified by an absorption model (TBABS; [17]) with free parameters of the column density ($N_{\rm H}$), the photon index ($\Gamma$) and the normalization. Then, we tried an absorbed SRCUT model instead of the PL model. SRCUT model describes synchrotron radiation from a PL distribution of electrons with an exponential cut-off [11]. The flux (at 1 GHz) for SRCUT is fixed at the value 1.24 Jy similar to previous studies [8, 9]. PL and SRCUT models are well fitted to the data. The parameters of both models for 2011 and 2015 observations are given in Table 1, where the uncertainties quoted are the 90\% confidence limits. As a next step, the SRCUT model was also applied with the spectral index fixed to 0.62 at 1 GHz as reported by [18]. This model also well reproduces the spectra with best-fit values shown in Table 1. XIS spectra for 2011 and 2015 observations are shown in Figure 2.

\begin{figure}
\centering \vspace*{1pt}
\includegraphics[width=0.49\textwidth]{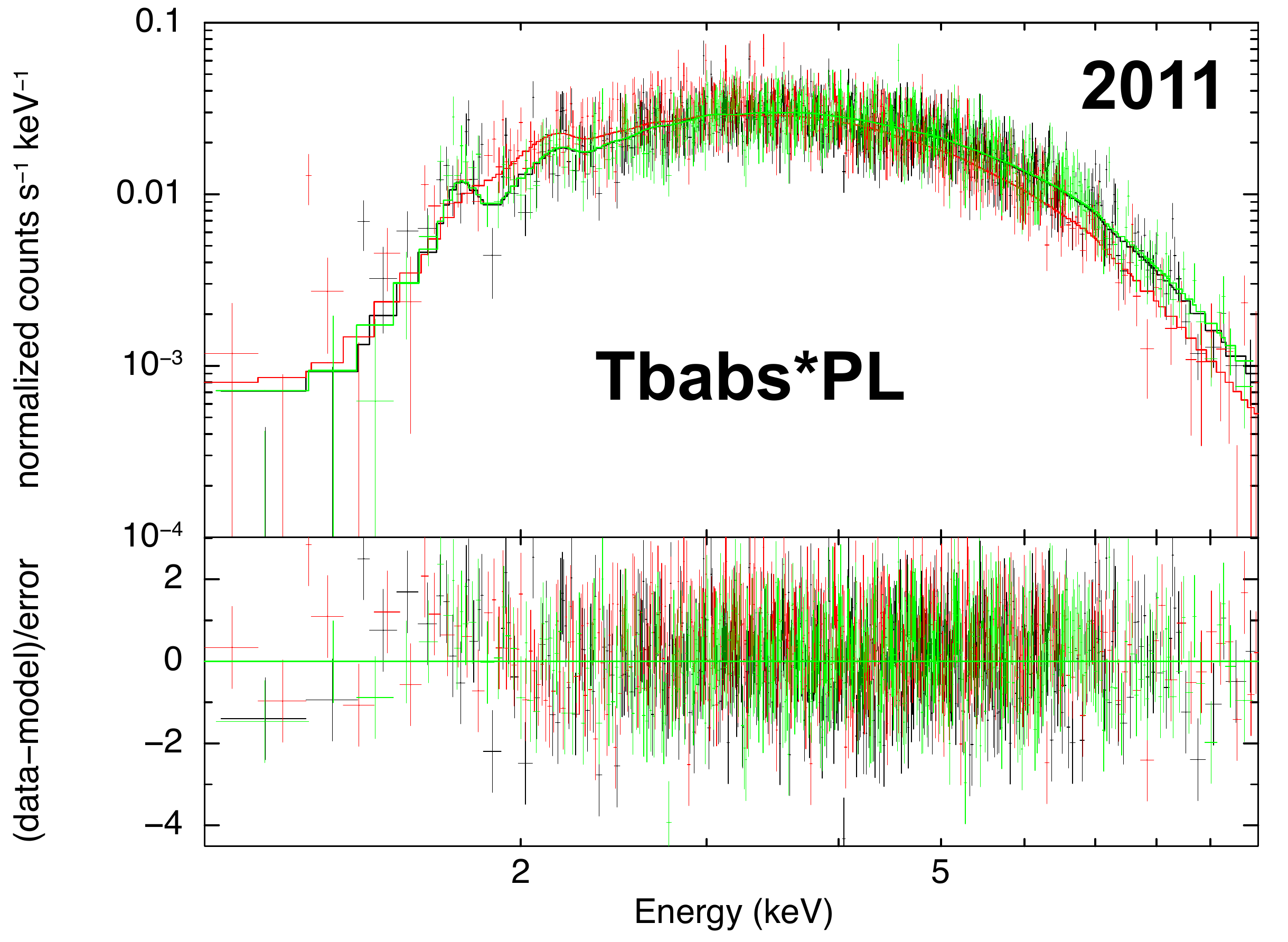}
\includegraphics[width=0.49\textwidth]{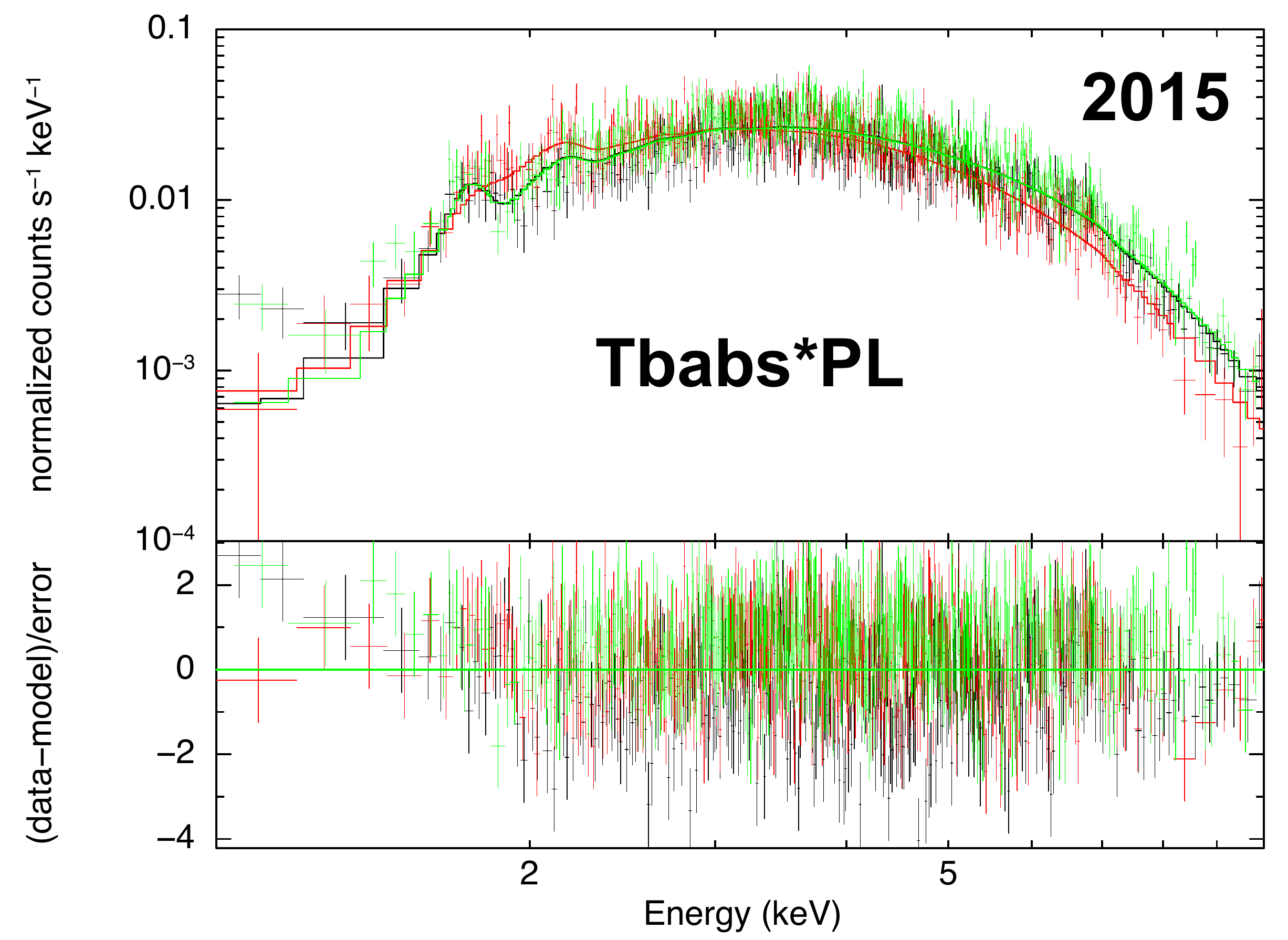}
\caption{Background-subtracted XIS0 (black), XIS1 (red) and XIS3 (green) spectra of G1.9+0.3.}
\label{figure_1}
\end{figure}

\section{Results and Discussion}
In this work, we focus on the time evolution of roll-off frequency of the synchrotron spectrum from SNR G1.9+0.3 using archival {\it Suzaku} data. We found that the roll-off frequencies of both observations are in the range of $\sim$($3.1-3.7$)$\times10^{17}$ Hz, which are consistent with the roll-off frequencies derived by [8, 9]. The roll-off frequency has been measured in several young SNRs (e.g., SN 1006 [19]). They found that the non-thermal component of X-ray spectra has a roll-off frequency of $\sim$$5.7\times10^{16}$ Hz, which is lower than the roll-off frequency found in our analysis.

From our spectral analysis, as seen Table 1, we find no significant differences in the spectral parameters between 2011 and 2015 observations. The roll-off frequency can be written as 
\begin{eqnarray}
{\nu_{\rm roll} \sim {\rm 0.97\times10^{17}} ~ {\rm Hz} ~ \xi^{-1} (V_{\rm s} / 10^8 ~ {\rm cm} ~ {\rm s}^{-1})^{2}}
\label{eqn:pi0}
\end{eqnarray}
where $\xi$ is the gyro-factor and $V_{\rm s}$ is the shock velocity (e.g., [20, 21]). The shock velocity of SNRs decreases with time after the free expansion phase [22]. Our results imply that the shock velocity is almost unchanged between 2011 and 2015. This may indicate that G1.9+0.3 is still in the free expansion phase.

%%% ACKNOWLEDGEMENTS
\section*{Acknowledgments}

AS is supported by the Scientific and Technological Research Council of Turkey (T\"{U}B\.{I}TAK) through the B\.{I}DEB-2219 fellowship program. This work is supported in part by grant-in-aid from the Ministry of Education, Culture, Sports, Science, and Technology (MEXT) of Japan, No.15K05088(RY), No.16K17702(YO), No.17K18270 (ST) and 16J06773(SK).

\begin{table*}
\caption{Results of Spectral Fitting}
 \begin{minipage}{170mm}
 \begin{center}
\begin{tabular}{@{}ccccc@{}}
    \hline
 \hline
&&\multicolumn{2}{c}{Observation}\\
\cline{3-4}
& Parameter  &  2011 &  2015\\
  \hline
&\multicolumn{3}{c}{TBABS*PL}\\
\hline
\\
& $N_{\rm H}$ ($\times10^{22}$ cm$^{-2}$)       &  $9.8^{+0.4}_{-0.4}$           &  $9.1^{+0.3}_{-0.3}$ &\\ [0.09 cm]
& Photon Index $({\Gamma})$                                   &  $2.52^{+0.07}_{-0.07}$        &  $2.49^{+0.06}_{-0.06}$ &\\ [0.09 cm]
%&& norm ($\times10^{-3}$ photon cm$^{-2}$ s$^{-1}$)   &  $6.7^{+0.9}_{-0.8}$           &  $5.7^{+0.7}_{-0.6}$ &\\ [0.09 cm]
& Flux$^{a}$ ($\times10^{-11}$ ergs cm$^{-2}$ s$^{-1}$) &  $2.33^{+0.11}_{-0.07}$           &  $2.01^{+0.09}_{-0.05}$ &\\ [0.09 cm]

& $\chi^{2}$/dof                                     &  1848.3/1801                   &  1596.5/1373 &\\
  \hline
&\multicolumn{3}{c}{TBABS*SRCUT}\\
\hline
  \\
& $N_{\rm H}$ ($\times10^{22}$ cm$^{-2}$)       &  $9.2^{+0.3}_{-0.4}$        &  $8.6^{+0.3}_{-0.2}$ & \\ [0.09 cm]
& $\alpha$                                      &  $0.608^{+0.013}_{-0.008}$     &  $0.616^{+0.011}_{-0.007}$ &\\ [0.09 cm]
& $\nu_{\rm rolloff}$ ($\times10^{17}$ Hz)           &  $3.12^{+1.14}_{-0.51}$           &  $3.33^{+1.04}_{-0.51}$  &\\ [0.09 cm]
& norm (Jy at 1 GHz)                                 &  $1.24$ (fixed)                &  $1.24$ (fixed) &\\ 
& $\chi^{2}$/dof                                     &  1842.5/1801                   &  1592.9/1373   &\\
 \hline
&\multicolumn{3}{c}{TBABS*SRCUT ($\alpha$=0.62)}\\
\hline
 \\
& $N_{\rm H}$ ($\times10^{22}$ cm$^{-2}$)  &  $8.9^{+0.2}_{-0.2}$        &  $8.6^{+0.3}_{-0.1}$ &\\ [0.09 cm]
& $\alpha$                                      &  $0.62$ (fixed)                     &  $0.62$ (fixed)      &   \\ [0.09 cm]
& $\nu_{\rm rolloff}$ ($\times10^{17}$ Hz)           &  $3.71^{+0.08}_{-0.09}$           &  $3.59^{+0.06}_{-0.09}$ & \\ [0.09 cm]
& norm (Jy at 1 GHz)                                 &  $1.24$ (fixed)                &  $1.24$ (fixed) &\\ 
& $\chi^{2}$/dof                                     &  1844.6/1802                   &  1603.4/1374  & \\
 \hline
\end{tabular}
\begin{flushleft}
$^{a}$ Unabsorbed fluxes in the 0.5$-$10.0 keV energy range. 
\end{flushleft}
\end{center}
\end{minipage}
\end{table*}

\section{References}
%%%%%%%%%%%%%%%%%%%%%%%%%%%%%%%%%%%%%%%%%%%

\smallskip

\end{document}